\documentstyle[twocolumn,prl,aps]{revtex}
\begin{document}
%\draft
\tighten
\title{The level shifting induced negative magnetoresistance 
in the nearest-neighbor hopping conduction}
\author{X. R. Wang and S. C. Ma}
\address{
Department of Physics, The Hong Kong University of Science
and Technology, Clear Water Bay, Hong Kong
}
\author{X. C. Xie}
\address{
Department of Physics, Oklahoma State University,
Stillwater, OK 74078
}
%\address{\rm (Submitted to Physical Review Letters in September 1997)}
\address{\rm (Draft on \today )}
\address{\mbox{ }}
%\address{\mbox{ }}
%\begin{abstract}
\address{\parbox{14cm}{\rm \mbox{ }\mbox{ }
We propose a new mechanism of negative magnetoresistance in 
non-magnetic granular materials in which electron transport is 
dominated by hopping between two nearest-neighbor clusters.
We study the dependence of magnetoresistance on 
temperature and separation between neighboring clusters.
At a small separation we find a negative magnetoresistance at low 
temperatures and it changes over to a positive value as 
temperature increases. For a fixed temperature, magnetoresistance changes
from negative to positive when the cluster separation increases.
The change of magnetoresistance $\Delta R/R$ can be more 
than 80\% at low temperatures. }}
\address{\mbox{ }}
%\address{\mbox{ }}
%\end{abstract}
\address{\parbox{14cm}{\rm PACS numbers: 72.80.Tm, 75.70.Pa, 81.05.Rm}}
\maketitle

Magnetoresistance in various transport mechanisms has been the 
subject of many studies. Negative magnetoresistance (MR)
has been observed and explained in many systems in the past
two decades\cite{plee}. Negative MR in dirty metals is related 
to weak localization phenomena\cite{plee}. Negative MR in Mott's 
variable-range hopping(VRH) region is also possible\cite{mksw} 
though backscattering is unimportant there, because electron 
states are highly localized and the probability of backscattering
is exponentially small. In the VRH case, it was pointed out\cite{mksw,sw}
that the ensemble average should be on the logarithm of conductance
so that interference\cite{mksw,sw,bottger} between forward tunneling paths
is important. The replica treatment\cite{sw} shows that pairing of 
tunneling paths is important to the average. A magnetic field 
can introduce phases to tunneling paths. Thus, the pairing is 
weaken, the localization length increases, and negative MR is resulted. 
Recently, there are reports that negative MR was observed in
Al/Al$_2$O$_3$\cite{abp} and gold\cite{belevtsev} granular 
materials in which the nearest neighbor hopping is the main transport 
mechanism. One of the interesting features is that negative MR  
occurs only near the percolation threshold\cite{abp,belevtsev}. It 
is unlikely that these experimental results can be explained either 
by the mechanism of band conduction or VRH conduction. In this 
letter, we propose a novel mechanism for negative MR in a granular 
material or a quantum dot array containing non-magnetic elements. 
We show that level shifting, magnetic confinement, and 
quantum interference in a 
phonon assisted hopping process can produce very rich physics. 

There are three energy scales which play important roles in a 
magneto-transport in granular metals and quantum-dot arrays. They 
are the thermal energy $kT$, the level spacing $\delta$, and the 
Zeeman energy. When the quantum dot sizes are small 
enough such that the typical level spacing is 
comparable to the thermal energy, only a few states near the 
Fermi energy contribute to hopping conduction. Therefore, if 
one can manipulate the relative level positions around the 
Fermi energy, one can then change the relative importance of these 
states to hopping conduction, which, in turn, may change 
magnetoresistance substantially. Depending on the relative direction 
between the electron magnetic moment and magnetic field, an external 
magnetic field can either increase or decrease the energy of an 
electron state through the Zeeman effect. 
This level shifting may enhance the phonon assisted hopping
and, thus, lead to a decrease of the resistance, namely, a 
negative MR in the nearest-neighbor hopping conduction. 
It is interesting to note that this mechanism of 
negative MR is completely different from that in 
Mott's variable-range hopping conduction\cite{mksw,sw}. 

In the dielectric regime of a granular metal, metallic clusters are 
embedded inside an insulating matrix\cite{ping}. Electrons reside 
on these metallic clusters, and they can move inside the material by 
hopping from one cluster onto another. Therefore, it is important to 
understand how an external magnetic field $B$ affects the electron 
hopping between two nearest-neighbor clusters. Consider a system 
with two clusters in the xy-plane centered at $(d/2,0)$ and 
$(-d/2,0)$, respectively. Each cluster is modeled by a parabolic 
confinement potential (This potential has been used to explain 
many experimental results\cite{pot}). 
The one-electron Hamiltonian is described by 
\begin{equation}
\label{hamiltonian}
H= -\frac{1}{2m}(\vec{p}-\frac{e}{c}\vec{A})^{2}+V(x,y)+\mu_{B}\sigma B,
\end{equation}
with 
\[
V(x,y)=\left\{ \begin {array}{ll} \frac{1}{2}m\omega_{r}^{2}(
(x-\frac{d}{2})^{2}+y^{2}) & x>0 \\  \frac{1}{2}m\omega_{l}^{2}(
(x+\frac{d}{2})^{2}+y^{2}) & x<0 \end{array} \right.
\]
where $m$ is the effective electron mass, and $\omega_r$ and $\omega_l$ 
are the simple harmonic oscillator frequencies parameterizing the 
right and the left metallic clusters, respectively. The last term in 
Hamiltonian (\ref{hamiltonian}) is the usual spin Zeeman interaction 
in which $\mu_{B}$ is the Bohr magneton and $\sigma=\pm 1$
correspond to spin-up and spin-down states, respectively. 
The larger $\omega$ is, the smaller a metallic cluster will be. 
When the two clusters are separated far away from each other, 
each cluster can be described well\cite{pot} by eigenenergies 
and eigenfunctions of a simple harmonic oscillator which can be 
analytically solved with eigenvalues 
$E_{n,j,\sigma}=(2n+|j|+1)\hbar \sqrt{\omega_{l,r}^{2}+\omega_{c}^{2}}
-j\hbar \omega_{c}+\mu_{B}\sigma B$, where $n=0,1,2,\ldots$, 
$j=0,\pm 1,\pm 2,\ldots$, $\sigma=\pm 1$ and $\omega_{c}=
Be/(2mc)$.
In the absence of a magnetic field, $\omega_{c}=0$, $E_{n,j}$ 
is $2(2n+|j|+1)$-fold degenerate. The degeneracies are broken 
due to the Zeeman effect. 

In the hopping regime, thermal energy $kT$ is much smaller 
than the Fermi energy $\mu$, and the two clusters are only 
weakly coupled. Hopping can be regarded as an  
electron jumping from a state $\psi_1$ in the left cluster to 
a state $\psi_2$ in the right cluster. In the tight-binding 
approximation, the tunneling matrix element is\cite{wx}
\begin{eqnarray}
\label{tun}
t_{12}=\frac{\hbar^2}{m}\int_{-\infty}^{\infty}&[&(
\psi_{1}^{\ast}\frac{\partial\psi_2}{\partial x}-
\psi_2\frac{\partial\psi_{1}^{\ast}}{\partial x})-\nonumber\\
&\frac{2i}{\phi_0}&(\vec{A}\cdot\hat{x})\psi_{1}^{\ast}\psi_2]\mid_{x=0}dy,
\end{eqnarray}
where $\phi_0=\frac{c\hbar}{e}$ is the flux quanta. In a uniform magnetic 
field $B$, with symmetric gauge $\vec A=(-By/2, Bx/2,0)$, the spatial 
parts of $\psi_1$ and $\psi_2$ can be expressed in terms of 
eigenstate $\psi_{n,j}$ of a simple harmonic oscillator as\cite{wx}
\begin{equation}
\label{leftwavefunction}
\psi_{1}=\exp (i\frac {e}{c\hbar}\vec{A_{0}}\cdot \vec{r}) 
\psi_{n,j}(r_{1},\theta_{1} ),
\end{equation}
\begin{equation}
\label{rightwavefunction}
\psi_{2}=\exp (-i\frac {e}{c\hbar}\vec{A_{0}}\cdot \vec{r}) 
\psi_{n,j}(r_{2},\theta_{2} ),
\end{equation}
where $\vec{A_{0}}=Bd\hat{y}/4$, $n$ and $j$ are the quantum numbers of 
a simple harmonic oscillators, $r_{1}$ and $\theta_{1}$ are the polar 
coordinates of $(r\cos\theta-d/2,r\sin\theta)$, and $r_{2}$ and 
$\theta_{2}$ are the polar coordinates of $(r\cos\theta+d/2,r\sin\theta)$.
The phase factors in Eqs. (\ref{leftwavefunction}) and 
(\ref{rightwavefunction}) are due to the magnetic field. They will 
give the usual interference on the tunneling matrix element. 
We will see later that this interference together with level shifting
and the usual magnetic confinement can induce both positive and 
negative magnetoresistance.

In a process that an electron loses completely its phase coherence
after it hops 
from one cluster onto another (incoherent hopping process), we can 
describe a granular system as a random resistor-network\cite{boris} 
in which each pair of metallic clusters is replaced by a resistor. 
If the electric field is small such that changes of the Fermi energy
and eigenenergies are negligible compared with $kT$, then 
resistance between two clusters can be calculated by\cite{boris}
\begin{eqnarray}
\label{resistance}
R\propto\sum_{l,r} \mid t_{lr}\mid^{-2}\exp(&-&\frac{1}{2kT}(\mid
\tilde{E}_l-\tilde{E}_r\mid+\mid\tilde{E}_l-\mu\mid+\nonumber\\
&\mid &\tilde{E}_r-\mu\mid)),
\end{eqnarray}
where $\mu$ is the Fermi energy. $\tilde{E}_l$ and 
$\tilde{E}_r$ are the energies of states $\psi_l$ and $\psi_r$,
respectively, including the contribution from electron-electron
interactions. The sum is over all states in both left and right 
clusters. Equation (\ref{resistance}) takes into account of a process 
that an electron in state $\tilde{E}_l$ ($\tilde{E}_r$) absorbs 
(emits) an phonon of energy $\tilde{E}_l - \tilde{E}_r$, and 
hops to unoccupied state $\tilde{E}_r$ ($\tilde{E}_l$).
The Fermi energy is related  
to the number of electrons $N$ in a cluster. At zero temperature, 
$\mu \sim \sqrt{N} \hbar \omega$, where $\omega$ is the harmonic 
oscillator frequency for the cluster. 
In the calculations shown below, we consider a process that 
one electron hops from one neutral cluster onto another neutral 
cluster. Electron-electron interactions are taken into account 
through the capacitor effect, namely, an extra charging energy, $e^{2}
/(\kappa r)$, will be added on every eigenenergy of single electron 
Hamiltonian (\ref{hamiltonian}) when an electron is put on a neutral 
cluster. $r$ is the size of the cluster. $\kappa$ is the effective 
dielectric constant. The Coulomb charging energy on a cluster of 
nano-meter can be much larger 
than the thermal energy. However, for an 
electron tunneling between two neighboring clusters of similar sizes,
the charging energy is close in value before and after tunneling process, thus, 
the Coulomb blockade effect is not significant. This is 
very much different from a Coulomb blockade system where electrons 
tunnel from a lead to a quantum dot\cite{Datta}.

\begin{figure}
%Fig1a
 \vbox to 6.8cm {\vss\hbox to 6.8cm
 {\hss\
   {\includegraphics{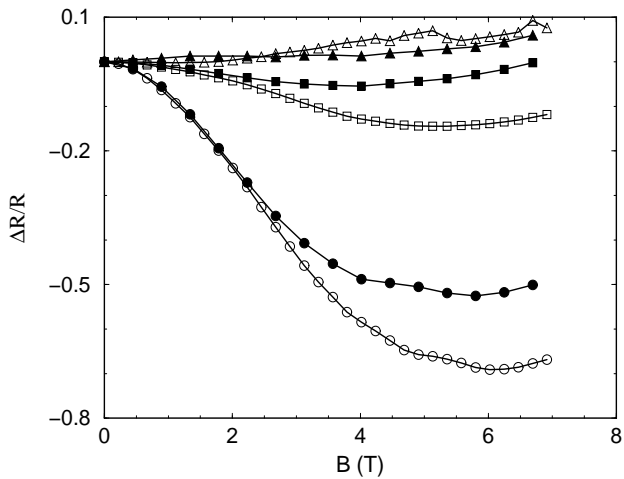}
   }
  \hss}
 } 
\vspace{6mm}
\caption{$\Delta R/R$ vs $B$ at different values of temperature
$T$. Open symbols are the data for a two-cluster system with
$\mu=11.75\hbar \omega_{0}$, $\omega_{l}=\omega_{0}=4.7$ meV/$\hbar$,
$\omega_{r}=0.965\omega_{0}=4.5$ meV/$\hbar$, and $d=80$ \AA.
(circle) $T=6$ K; (square) $T=12$ K; (up triangle) $T=30$ K.
Filled symbols are the data for a $10\times 10$ cluster array. 
The harmonic oscillator frequency of each cluster is random with 
Gaussian distribution function of mean $\omega_0$ and the mean 
square deviation $0.1\omega_0$. $\mu=11.75\hbar \omega_{0}$, 
$\omega_{0}=4.7$ meV/$\hbar$, and $d=80$ \AA. (circle) $T=6$ K;
(square) $T=12$ K; (up triangle) $T=18$ K.
}\label{fig1}
\end{figure}

Given $\omega_l$, $\omega_r$, the effective mass m, and the magnetic 
field $B$, it is straight forward to compute $R(B)$ numerically 
using Eq.(\ref{resistance}). To obtain 
some realistic numbers, we use, throughout this study, 
$m=1.1m_e$, the effective electron mass of a gold metal.
We assume the atom-atom distance to be 
$1.6$ \AA \ which is suitable for a gold metal. Let us 
consider a system with the Fermi energy $\mu=11.75\hbar \omega_0$. 
This corresponds to about 144 electrons on a cluster. The diameter 
of the cluster is then about $D=38.4$ \AA. We choose $\omega_0$
in such a way that $\hbar \omega_0$ is equal to the typical level 
spacing of this kind of clusters, i.e. $\hbar \omega_0 =\hbar^{2}/(
mD^{2})=4.7$ meV. In order to eliminate possible artificial 
effects such as resonance tunneling, we assume that the two 
clusters are not identical by choosing $\omega_{l}=4.7$ meV/$\hbar$ 
and $\omega_{r}=4.5$ meV/$\hbar$. The open symbols in Figure \ref{fig1}
are numerical results of the dimensionless MR 
$\Delta R/R=(R(B)-R(0))/R(0)$ vs $B$ when the cluster-cluster 
separation is about $d=80$ \AA. At temperature $T=6$ K, the 
MR is negative. It starts to saturate around 
magnetic field $B=6$ T. The change in resistance is about 
$70\%$, comparable to giant magnetoresistance (GMR) obtained 
in magnetic layer systems\cite{chien,xjc}. At $T=12$ K, 
MR is still negative, but the resistance change 
is substantially reduced to about $10\% $. At $T=30$ K, the 
MR becomes positive, and the resistance increases 
by about $10\%$ at magnetic field around $7$ T. These results can 
be understood from the following arguments. At very low temperatures 
($KT\ll \hbar \omega_{l}, \ \hbar \omega_{r}$), only those states 
near the Fermi level, which lies in general between two discreted 
levels of a cluster, are important to the electron transport. 
In the presence of a magnetic field, at least two of these states 
approach to the Fermi level (one from below and one from above),
due to the breaking of time-reversal symmetry through the orbital Zeeman 
effect. According to Eq.(\ref{resistance}), resistance will be 
reduced because of hopping between the two states. With 
increase of temperature, more states will play significant 
roles in hopping conduction. The energy shifting toward the 
Fermi energy becomes less important, and the phase interference 
on the tunneling matrix element (Eq.(\ref{tun})) dominates. 
This leads eventually to a positive MR 
at high temperature when the magnetic field is strong.

The behavior of a granular material is very similar to that of 
a two-cluster system. We also carried out a similar calculations 
on a $10\times 10$ array system with the Fermi energy 
$\mu=11.75\hbar \omega_0$. Each cluster is still 
modeled by a harmonic potential with a random $\omega$. $\omega$ 
has a gaussian distribution with the mean value $\omega_{0}$ 
and the mean square deviation $0.1\omega_{0}$. The 
separation between two nearest-neighbor clusters is $80$ \AA. 
The results of $\Delta R/R=(R(B)-R(0))/R(0)$ vs $B$ for various
temperatures are shown in Figure \ref{fig1} by using filled symbols.
The behavior is very similar to that of the case with two clusters.

\begin{figure}
%Fig1a
 \vbox to 6.5cm {\vss\hbox to 6.5cm
 {\hss\
   {\includegraphics{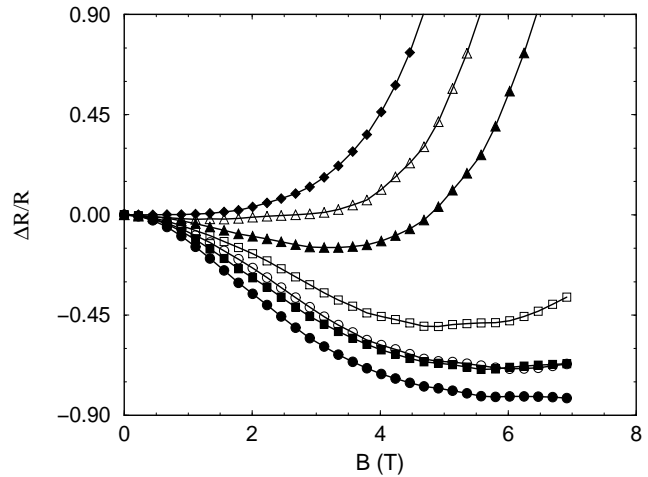}
   }
  \hss}
 }
\vspace{6mm}
\caption{$\Delta R/R$ vs $B$ at $T=6$ K for different values of 
cluster-cluster separation $d$. Open symbols are the data for 
a two-cluster system with $\mu=11.75\hbar \omega_{0}$,
$\omega_{l}=\omega_{0}=4.7$ meV/$\hbar$, and 
$\omega_{r}=0.965\omega_{0}=4.5$ meV/$\hbar$. (circle) $d=80$ \AA; 
(square) $d=107$ \AA; (up triangle) $d=160$ \AA. 
Filled symbols are the data for a two-cluster system with
$\mu=11.35\hbar \omega_{0}$, $\omega_{l}=\omega_{0}=4.7$ meV/$\hbar$,
and $\omega_{r}=0.965\omega_{0}=4.5$ meV/$\hbar$. 
(circle) $d=80$ \AA; (square) $d=107$ \AA; (up triangle) $d=160$ \AA; 
(diamond) $d=187$ \AA.
}\label{fig2}
\end{figure}

According to the phase factors in Eqs. (\ref{leftwavefunction}) and 
(\ref{rightwavefunction}), one expects that interference on 
tunneling matrix element $t_{12}$ should increase with 
cluster-cluster separation $d$. We test this
by studying a two-cluster system of the Fermi energy $\mu=11.75\hbar 
\omega_{0}$, $\omega_{l}=\omega_{0}=4.7$ meV/$\hbar$, and
$\omega_{r}=0.965\omega_{0}=4.5$ meV/$\hbar$. The 
dimensionless MR $\Delta R/R=(R(B)-R(0))/R(0)$ vs $B$ with 
various $d$ at temperature $T=6$ K is calculated. The numerical 
results are shown by the open symbols in Figure \ref{fig2}. 
It shows clearly that MR changes from negative
sign to positive sign as the separation increases from 
$d=80$ \AA \ to $d=160$ \AA.

The general behavior of MR should be robust to a
particular choice of the Fermi level. To demonstrate this, we 
calculate $R(B)$ for a different value of the Fermi energy 
$\mu=11.35\hbar \omega_0$, but with the same $\omega_l$ and 
$\omega_r$. Filled symbols in Figure \ref{fig2} are $\Delta 
R/R=(R(B)-R(0))/R(0)$ vs $B$ at $T=6 $K with cluster-cluster 
separation $d=80,\ 107,\ 160,\ 187$ \AA. Comparing the results 
with those of the Fermi energy $\mu=11.75\hbar \omega_0$, we 
see indeed that the qualitative features remain the same. 
MR undergoes a sign transition as the 
cluster separation increases from $d=80$ \AA \ to $d=187$ \AA.

It is worth to stress
that the nearest neighbor hopping conduction in homogeneous materials
leads usually to a positive magnetoresistance\cite{boris}
due to the destructive quantum interference and magnetic
confinement. However,
we have used a simple model to demonstrate that the level shifting
can lead to a negative magnetoresistance in the nearest-neighbor
hopping conduction in granular materials. 
The negative MR was observed in recent experiments on 
Al/Al$_2$O$_3$\cite{abp} and gold granular materials\cite{belevtsev}. 
Our theory predicts a sign change of MR from negative to positive as 
temperature increases, and a similar sign change was also seen 
experimentally\cite{belevtsev}. According to the theory, a sign 
change should also occur as cluster-cluster distance increases 
since then the quantum interference and magnetic confinement will 
play more important roles in the hopping process. This is consistent 
with the experiments where the negative MR was only observed near 
the metal-insulator transition (near the percolation threshold) 
and on the insulator side\cite{abp,belevtsev}. 

Our theory suggests a new type of non-magnetic 
materials which can produce a GMR ($>80\% $). Unlike the usual 
magnetic systems in which the change of spin-related scattering in 
a magnetic field is responsible to the observed GMR\cite{chien,xjc}, 
the mechanism here is the level shifting due to the Zeeman 
interaction. There is no hysteresis in the system since it is 
non-magnetic in nature. This property might be an advantage in 
real applications. 

We have used a symmetric confinement potential to model the quantum 
dots. The spherical symmetry gives rise to a degeneracy beside the spin
degeneracy at zero magnetic field which will guarantee that there 
are always some states moving closer to the Fermi energy as magnetic
field increases. Closing of energy levels will enhance the phonon 
assistant hopping which, in turn, leads to a large negative MR. 
However, slight asymmetry will not destroy the above mechanism.
In a system with asymmetric dots, the spin degeneracy survives, and 
the level closing will be gradual.
Thus, the effects discussed in this work, such as GMR and sign change
of MG will occur at a lower temperature. On the other hand,
in a real granular metal, metallic clusters tend to form a 
spherical shape due to a large surface tension. Therefore, a 
metal with a large surface tension in an insulating matrix is 
a good candidate to observe the transport property discussed here.
Also, one expects that the negative MR can be enhanced by annealing 
a sample because this process can make metallic clusters more 
symmetric. The nano-technology allows us to make artificial atoms,
molecules, and quantum dot arrays via surface gates in heterostructures.
It has been showed that quantum dots in such systems can be 
well described by symmetrical potentials\cite{pot,kumar}. 
This may be an ideal system to test our theory with 
tunable dot size and dot-dot separation. Further studies along 
this direction should be interesting.
Although we only studied a specific 2D model, but the physics 
discussed in this paper is expected to carry over to other 2D as 
well as 3D models. 

In summary, we propose a novel mechanism of GMR in
non-magnetic granular materials in which the main
transport process is hopping between
nearest-neighbor clusters. We find that 
magnetoresistance changes sign as the temperature
or separation between clusters varies. The unexpected negative 
MR is due to the level shifting of the Zeeman effect.

It is our pleasure to thank Ping Sheng for many useful comments and 
suggestions, and for his drawing our attention to the experimental 
results of references 5 and 6. We also acknowledge Q. Niu for helpful 
discussions and B. I. Belevtsev for informing us their experimental 
results prior to publication. 
This work is supported in part by UGC, Hong Kong, 
through DAG grants. X. C. Xie is supported by DOE.

% now the references. delete or change fake bibitem. delete next three
%   lines and directly read in your .bbl file if you use bibtex.


\begin{references}
\bibitem{plee} G. Bergmann, Physics Reports {\bf 107}, 2 (1984); P. A. Lee
and T. V. Ramakrishnan, Rev. of Mod. Phys. {\bf 57}, 287 (1985).
\bibitem{mksw} V. L. Nguyen, B. Z. Spivak and B. I. Shklovskii, Sov. 
Phys. JETP Lett. {\bf 41} 42(1985);
V. Sivan, O. Entin-Wohlman, and Y. Imry,
Phys. Rev. Lett. {\bf 60} 1566(1988); O. Faran and Z. Ovadyahu,
Phys. Rev. {\bf B38} 5457(1988).
\bibitem{sw} Y. Shapir, X. R. Wang, Europhys. Lett. {\bf 4}, 1165(1987);
E. Medina, M. Kardar, Y. Shapir, and X. R. Wang, Phys. Rev.
Lett. {\bf 62}, 941 (1989); {\it ibid.} {\bf 64}, 1816(1990); X. R. Wang,
Y. Shapir, E. Medina, and  M. Kardar, Phys. Rev. B. {\bf 42}, 4559(1990); 
\bibitem{bottger} H. B\"ottger, V. V. Bryksin, and F. Schulz, Phys. Rev. B
{\bf 49}, 2447(1994).
\bibitem{abp} A. B. Pakhomov, D. S. McLachlan, I. I. Oblakova, and 
A. M. Virnik, J. Phys.: Condens. Matter {\bf 5}, 5313 (1993). 
\bibitem{belevtsev}B. I. Belevtsev, E. Yu. Beliayev, and E. Yu. 
Kopeichenko, Low Temperature
Physics 23, 724(1997).
\bibitem{ping} P. Sheng, Philo. Mag. B {\bf 65}, 357(1992); B. Abeles, 
P. Sheng, M. D. Couttis, and Y. Arie, Adv. Phys. {\bf 24}, 407(1975).
\bibitem{pot}T. Schmidt et al., Phys. Rev. Lett. {\bf 78}, 1544 (1997); 
L. P. Kouwenhoven et al., Science {\bf 278}, 1788 (1997).
\bibitem{wx} X. R. Wang and X. C. Xie, Europhys. Lett. {\bf 38}, 55(1997).
\bibitem{boris} B. I. Shklovskii and A. L. Efros, in {\it Electronic
Properties of Doped Semiconductors} (Springer, Berlin, 1984);
H. B\"ottger and V. V. Bryksin, {\it Hopping Conduction in
Solids} (VCH Publisher, 1985).
\bibitem{Datta} S. Datta, {\it Electronic Transport in Mesoscopic
Systems} (Cambridge University Press, 1995).
\bibitem{chien} M. N. Baibich et al., 
Phys. Rev. Lett. {\bf 61}, 2472 (1988); 
P. Xiong et al., Phys. Rev. Lett. {\bf 69}, 3220 (1992).
\bibitem{xjc} J. Q. Xiao, J. S. Jiang, and C. L. Chien, 
Phys. Rev. Lett. {\bf 68}, 3749 (1992).
\bibitem{kumar} A. Kumar, S. Laux, and F. Stern, Phys. Rev. B {\bf 42},
5166 (1990); M. Stopa, Phys. Rev. B {\bf 54}, 13767 (1996).
\end{references}
\end{document}